# Suppression of ion conductance by electro-osmotic flow in nano-channels with weakly overlapping electrical double layers

Yang Liu[1*], Lingzi Guo[1], Xin Zhu[1], Qiushi Ran[2], Robert Dutton[2]

[1]College of Information Science and Electronic Engineering

Zhejiang University, Hangzhou, China

[2]Department of Electrical Engineering

Stanford University, Stanford, CA U.S.A.

**Abstract:**

This theoretical study investigates the nonlinear ionic current-voltage characteristics of nano-channels that have weakly overlapping electrical double layers. Numerical simulations as well as a 1-D mathematical model are developed to reveal that the electro-osmotic flow (EOF) interplays with the concentration-polarization process and depletes the ion concentration inside the channels, thus significantly suppressing the channel conductance. The conductance may be restored at high electrical biases in the presence of recirculating vortices within the channels. As a result of the EOF-driven ion depletion, a limiting-conductance behavior is identified, which is intrinsically different from the classical limiting-current behavior.

## 1. Introduction

Nano-fluidic channels have important applications in membrane technologies [1-3], analytical sample preparation [4-8], current rectification [9-11], and field-effect gating [12-15]. Understanding the ionic current-voltage characteristics in such devices has been a focus of extensive research efforts. The majority of the works pertain to channels that have strong overlap between the electrical double layers (EDLs), i.e. the channel height is comparable to or even smaller than the Debye screening length, $\Lambda_D$. In such a classical regime, a remarkable electrical characteristics is the limiting current behavior, where the ionic current approaches a limiting value at elevated voltage biases; its cause is well-understood as from the concentration polarization (CP) process [2,4,16-19]. There are other effects that are still under active study, such as the extended space charge layers, the vortex formation and their relation to overlimiting currents [5,20-25]. In general, the electro-osmotic flow (EOF) inside the nano-channels does not play a significant role in the classical regime [26].

More recently, nano-channels with weak EDL overlap have attracted great research interest for their unique device characteristics and relaxed constraints on fabrication [10,13,27-31]. In this new regime, both unipolar and ambipolar ion transport processes coexist; the EOF inside the nano-channels can be significant and has been experimentally used for DNA translocation modulation [13,32], current rectification [31], and enhanced molecular binding [33]. Its impact on the nonlinearity of ionic currents has also been briefly discussed in a numerical study for nanopores that are voltage-gated by embedded side electrodes [27,28]. In a previous work by Mani et al. [6], the interplay of the EOF and CP processes at the interface of micro-channel and nano-channel has been studied comprehensively; it has been particularly shown that, under the non-propagation CP condition, the ion concentration inside the nano-channel is suppressed by the localized CP effect. In the present work, we aim to numerically examine the nonlinear current-voltage characteristics in nano-channels with weak EDL overlap and reveal a limiting-conductance behavior that is intrinsically different from the limiting-current behavior classical regime. It will be shown that this unique characteristics can be explained by the general theory of Mani et al.[6]

Furthermore, a refined 1-D mathematical model will be developed to give a simple yet accurate account of the limiting conductance.

**2. Results and Discussions**

A basic nano-channel structure as schematically shown in Fig. 1(a) is studied without the loss of generality. The cation-selective nano-channel has a fixed surface charge density, $-\sigma$, and connects two micro-reservoirs filled with KCl solution of a bulk ion concentration, $C_0$. Only the top-half of the channel is modeled for symmetry consideration, and the y-dimension is assumed infinite. An electrical bias, $V_d$, is applied between the two reservoirs and generates an ion current, $I_d$. Here, the EOF flows from left to right, thus defining the channel entrance and exit, respectively. Our study pertains to the regime of weak EDL overlap, i.e. $h \geq 5\Lambda_D$, where $h$ is the channel half-height.

It is noted that there are two common types of nano-fluidic channels[12]: 3D nanopores/nano-tubes and 2D nano-slits. This study models the nano-slit structures by assuming that the channel width is much greater than the channel height. Nonetheless, the nonlinear conductance effect revealed in this paper is rather general and also occurs in cylindrical nano-pore structures. Additional simulation results for a cylindrical nanopore structure with weak EDL overlap are presented in the Supplemental Material.[35]

Numerical simulations are conducted using both the Poisson-Nernst-Planck (PNP) [16] and Poisson-Nernst-Planck-Stokes (PNP-S) [27] models. The former neglects any fluid transport while the latter models it as low-Reynolds-number flow with the hydrophilic, no-slip boundary condition [34]. In this study we use a finite-volume device simulator, PROPHET, and the simulations are verified by another finite-element one, COMSOL, to rule out possible numerical artifacts [35].

The simulated current-voltage curves are shown in Fig. 1(b) for $h = 50nm$, $L = 1\mu m$, $H' = 1.05\mu m$, $L' = 1\mu m$. The bulk ion concentration $C_0$ is 1mM, corresponding to a $\Lambda_D$ of 10nm. The surface charge density, $-\sigma$, is set to a typical value of $-0.01q/nm^2$. For the PNP model, the differential conductance slightly decreases with increasing $V_d$. This is expected from the classical limiting-current theory: the moderate ion selectivity from the weak EDL overlap results in a weak limiting-current behavior [16]. In contrast, the PNP-S model produces a strong nonlinear I-V curve: a severe decrease in the differential conductance is observed at moderate $V_d$ biases; as $V_d$ increases further ($> \sim 8V$ in this particular device), the differential conductance rapidly restores to a high value that is even greater than that of PNP. To further illustrate this effect, PNP-S simulations are carried out by artificially increasing the fluid viscosity from its nominal value $\eta_0 = 10^{-3} Ns/m^2$ to $2\eta_0$ and $10\eta_0$ (Fig. 1(b)). As the fluid flow is suppressed by the viscosity increase, the I-V curves become less nonlinear and approach to that of PNP.

The simulated fluid flow patterns in the channel region are plotted in Fig. 2 for two specific $V_d$ values representing the conductance suppression (5V) and restoration (12V) stages, respectively. Inside the entrance reservoir, a vortex can be seen for both biases, and the recirculating magnitude increases from 5V to 12V. This type of vortices is known for sharp EOF transition from the micro-reservoir to the nano-channel [6,26]. The most remarkable difference between the two biases is the flow pattern change inside the channel. At 5V, the flow lines exhibit a typical EOF velocity profile that is parallel to the channel surface. In contrast, a recirculating vortex is generated inside the channel at 12V.

The impact of the flow pattern on ion concentration is also examined in Fig. 2. Here, we define the mean ion concentration $C_m \equiv (C_+ + C_-)/2$, where $C_+$ and $C_-$ are the cation and anion concentrations, respectively. The plotted quantity $\Delta C_m$ is obtained by subtracting the $C_m$ distribution of PNP from its PNP-S counterpart. At 5V, the ion concentration is significantly depleted inside the channel. In contrast, the change is small at 12V. These observations confirm that the conductance suppression and restoration are intrinsically related to the fluid patterns.

The effect of ion depletion inside the nano-channel has been accounted for in the general theory on CP propagation by Mani et al.[6] Qualitatively, it can be understood by considering the interplay of the CP and EOF processes. The classical theory of CP does not account for EOF and states that the ion depletion occurs inside the entrance reservoir[19]. For channels with weak EDL overlap studied in this work, however, there exists an ambipolar portion (where $C_+ \sim C_-$ by definition) of the channel that connects the reservoirs, as schematically shown in the inset of Fig. 3(a). The ambipolar ion transport is of the convection-diffusion type: the EOF can drive the ion depletion zone from the reservoir into the channel and thus dramatically suppresses its conductance. This depletion process is alleviated when the vortex occurs to mix the ion solution inside the channel.

To further quantitatively analyze the EOF-induced ion depletion process, a 1-D model is developed for long channels following the approach by Dydek et al.[21], in which the PNP-S equations are averaged over the transversal direction. It is noted that, in the work of Mani et al.[6], a 1-D model has been developed to account for the ion transport in an analytical approach, but the EDL was simplified as a delta distribution of counter-ions. Here, we use a refined model to take into account the realistic ion distribution normal to the channel walls. This model refinement is important to achieve a quantitative agreement with the full numerical simulations.

As a benchmark, numerical simulations are performed for a long-channel structure with $h = 100 nm$, $L = 10 \mu m$, $H' = 1.1 \mu m$, $L' = 10 \mu m$, $-\sigma = -0.02 q/nm^2$, and $C_0 = 1 mM$ (Fig. 3a). Note that only the conductance suppression stage is present for the bias range in this case. The simulated profiles of ion concentration along the center ($x = 0$) at varying biases are plotted for PNP (Fig. 3b) and PNP-S (Fig. 3c) models, respectively. In the former, the simulated concentration polarization is in agreement with the classical theory [16]. Inside the channel, the ions are accumulated in most part and only depleted near the entrance. In the latter, drastically different profiles are observed due to EOF: the ion depletion zone shifts from the entrance reservoir into the channel.

The 1D model along the longitudinal direction is expressed for the channel portion ($0 \leq z \leq L$) as

$$-D\,dC_+(z)/dz + \mu C_+(z)E_z(z) + C_-(z)u_z + \tfrac{1}{2}\Sigma u_z = f_+, \qquad \text{Eq. (1)}$$

$$-D\,dC_-(z)/dz - \mu C_-(z)E_z(z) + C_-(z)u_z = f_-, \qquad \text{Eq. (2)}$$

$$C_+(z) = C_-(z) + \Sigma. \qquad \text{Eq. (3)}$$

Here, $C_+$, $C_-$, and $E_z$ are the cation concentration, anion concentration, and electric field, respectively. They are all averaged over the transversal direction, i.e. $C_+(z) \equiv 1/h \int C_+(x,z)dx$, and so on, based on the assumption $h \gg \Lambda_D$. The parameter $D$ is the ion diffusivity, and $\mu$ the ion mobility. The average ion fluxes, $f_+$ and $f_-$, remain constant along the channel due to continuity. The surface charge is also converted to an effective volume concentration, $\Sigma \equiv \sigma/h$. The EOF slip velocity $u_z$ is assumed constant along the channel. In Eq. (1), the cation convection term, $1/h \int C_+(x,z) u_z(x)dx \approx C_-(z)u_z + \tfrac{1}{2}\Sigma u_z$, accounts for both the ambipolar and surface EDL transport [35]. An auxiliary relation is also valid for tall channels [35]

$$C_\pm(z) \approx \pm \Sigma/2 + \sqrt{(\Sigma/2)^2 + C_m(z)^2}, \qquad \text{Eq. (4)}$$

where $C_m(z)$ is the ambipolar ion concentration along $x = 0$.

The case without EOF is firstly examined by letting $u_z = 0$, and a simple analytical expression for the $C_-$ and $z$ relation is obtained

$$-\frac{f_+ + f_-}{2D} z = Q(C_-) - Q(C_{-,0}), \qquad \text{Eq. (5)}$$

where the function $Q(C) \equiv \frac{\Sigma}{2}\frac{f_+ - f_-}{f_+ + f_-} \ln\left(C + \frac{f_-}{f_+ + f_-}\Sigma\right) + C$ is defined, and $C_{-,0}$ is the boundary value at $z = 0$. The derivation steps are given in the Supplemental Material [35]. Subsequently, $C_+$ and $C_m$ can be readily obtained from Eq. (3) and (4). In principle, the parameter values, $C_{-,0}$, $f_+$, and $f_-$ can be determined by coupling the model to that of the reservoirs. Here, we are mostly concerned about the functional form of $C_-(z)$. These parameters are therefore extracted as input from numerical simulations [35].

In the general case when the EOF velocity $u_z$ is non-negligible, it is obtained

$$\frac{2u_z}{D}(z - L) = R(C_-) - R(C_{-,L}),$$  Eq. (6)

where $R(C) \equiv \left(1 + \frac{\Sigma u_z + f_d}{\sqrt{\Delta}}\right) ln(4u_z C + \Sigma u_z - f_d - \sqrt{\Delta}) + \left(1 - \frac{\Sigma u_z + f_d}{\sqrt{\Delta}}\right) ln(4u_z C + \Sigma u_z - f_d + \Delta)$, and $C_{-,L}$ is the boundary value at $z=L$. The quantities $f_d \equiv f_+ + f_- - 1/2\,\Sigma u_z$ and $\Delta \equiv (\Sigma u_z - f_d)^2 + 8f_-\Sigma u_z$ are defined for convenience. The additional input parameter $u_z$ is also extracted from simulations. The derivation of Eq. (6) can be found in the Supplemental Material.[35]

In Figs. 3(d) and 3(e), the profiles of $C_m(z)$ calculated from Eqs. (5) and (6) are plotted for the cases without and with the EOF, respectively. The results show good agreement with simulations. It clearly demonstrates that the EOF is responsible for the drastic difference in the ion distribution functions.

A further inspection of Fig. 3(c) reveals a remarkable asymptotic behavior: as the bias increases, the EOF-driven ion depletion makes the channel concentration approach a limiting constant value. At the same time, the ion concentration in the reservoirs approaches $C_0$. Such a limiting behavior agrees with the scenario of non-propagating CP as revealed in the work of Mani et al.[6] In this limiting condition, the diffusion is negligible compared to the drift and convection processes. Balancing the total fluxes of the reservoirs and channel leads to rate equations:

$$\begin{cases} \mu H' C_0 E_z{'} + H' C_0 u_z{'} = \mu h(C_-^* + \Sigma) E_z + h C_-^* u_z + 1/2\, h\Sigma u_z \\ \mu H' C_0 E_z{'} - H' C_0 u_z{'} = \mu h C_-^* E_z - h C_-^* u_z \end{cases},$$

where $C_-^*$ is the limiting anion concentration, and $E_z, E_z{'}, u_z, u_z{'}$ are the electric field and fluid velocities in the channel and reservoirs, respectively. The charge neutrality, $C_+^* = C_-^* + \Sigma$, is implicitly assumed. By further applying the Helmholtz-Smoluchowski relation $u_z = \frac{\varepsilon_w \psi_s}{\eta} E_z$, the expression for the surface potential, $\psi_s = 2V_t \sinh^{-1}\left[\frac{\Sigma}{(C_r)^{1/2}(C_-^{*2}+\Sigma C_-^*)^{1/4}}\right]$ [35], and the connecting condition, $H'u_z' \approx (h - \Lambda_D)u_z$, a self-contained expression for $C_-^*$ is obtained

$$C_-^* \approx \left(1 - \Lambda_D/h\right)C_0 - \Sigma/4 - \frac{\eta\mu\Sigma}{4\varepsilon_w V_t \sinh^{-1}\left[\frac{\Sigma}{(C_r)^{1/2}(C_-^{*2}+\Sigma C_-^*)^{1/4}}\right]}, \quad \text{Eq. (7)}$$

where $V_t \equiv k_B T/q$ is the thermal voltage, $\varepsilon_w$ the water permittivity, and $C_r \equiv 8\varepsilon_w V_t/qh^2$.

The $C_-^*$ values calculated from Eq. (7) are plotted against the surface charge $\Sigma$ in Fig. 4 and show good agreement with those extracted from the PNP-S simulations. As $\Sigma$ increases, $C_-^*$ decreases toward zero, implying an increase in $\Lambda_D$. At very high $\Sigma$ values, the approximation $\Lambda_D \ll h$ no longer holds, causing the observed deviation in that end. Another disparity occurs as $\Sigma$ approaches zero. The 1D model assumes sufficiently strong EOF so that the limiting condition is reached. This would require infinitely high $V_d$ as $\Sigma$ approaches zero.

In the limiting condition, the overall conductance is expressed as

$$G^* = \left(1/G_C^* + 1/G_R^*\right)^{-1}, \quad \text{Eq. (8)}$$

where the channel conductance is $G_C^* \approx \left[\mu(2C_-^* + \Sigma) + \varepsilon_w\psi_s/2\eta \Sigma\right]qh/L$, and the reservoir conductance is $G_R^* \approx q\mu C_0 H'/L'$. In Fig. 5(a), we re-plot the simulated I-V curve of the PNP-S model from Fig. 3(a). The asymptote $I_d = G^* V_d$ is plotted using the limiting conductance calculated from Eq. (8). As a reference, another asymptote $I_d = G_0 V_d$ is plotted using the classical expression, $2\sqrt{(\Sigma/2)^2 + C_0^2}\,\mu qh/L$, for the channel conductance at low biases[26]. Clearly, the ion conductance approaches the predicted limiting value at high biases. More data of the limiting conductance are plotted in Fig. 5(b) for varying surface charge density. Quantitative agreement between the calculations and simulations is observed, except for the disparity at the zero bias as explained above.

We emphasize that the quantity approaching a limiting value here is the conductance $I_d/V_d$, as clearly shown in Fig. 5(a), rather than the current $I_d$ as commonly studied [26]. The limiting-current and limiting-conductance processes are both related to the concentration polarization, but they are intrinsically different. In the former, the ion depletion occurs in the entrance reservoir, and the diffusion therein limits the overall conductance. In the latter, the ion depletion is driven into the channel, which becomes the conductance bottleneck.

Our analysis is focused on the ion depletion process, since it is the cause of the nonlinearity in the first place. A quantitative model on the vortex generation and conductance restoration is beyond the scope of this study. We instead note that the over-limiting current and associated vortices (commonly referred to those generated in the reservoirs) are still actively studied from aspects such as hydrodynamic instability[23,24] and EOF back-flow[21]. It is also known that the variation of the channel height can lead to vortex generation both inside the channel and at the openings[8]. In the present study, there exists a variation in the screening length, particularly near the channel openings. Whether the channel vortex observed in this study shares one of those generation mechanisms or has a different origin remains to be further investigated.

## 3. Conclusions

In summary, the nonlinear current-voltage characteristics of nano-channels with weak EDL overlap are numerically studied. It is shown that the EOF drives the ion depletion zone into the channels and suppresses the ion conductance. The conductance may be restored at high electrical biases due to the occurrence of vortices inside the channels. It is further revealed that, in the conductance suppression stage, the I-V characteristics exhibit a limiting-conductance behavior that is intrinsically different from the classical limiting-current behavior. This unique limiting behavior is explained by the fact that the depleted ion concentration asymptotically approaches a limiting value. A simple algebraic equation is established to calculate the limiting conductance.

Y.L., L.G., and X.Z. acknowledge the support of National Natural Science Foundation of China (Grant No: 61574126) and Natural Science Foundation of Zhejiang Province, China (Grant No: LR15F040001).

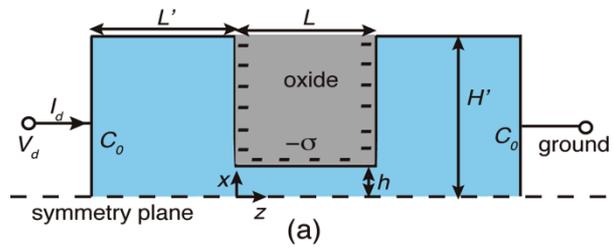

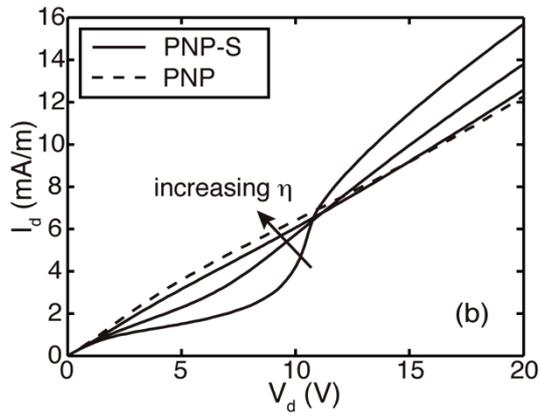

*Fig. 1. (a) Schemmatic of a nano-channel structure; (b) Simulated current-voltage curves using both the PNP and PNP-S models. For the PNP-S model, three different viscosity values, $\eta_0 = 0.001 Ns/m^2$, $2\eta_0$, and $10\eta_0$, are simulated, respectively.*

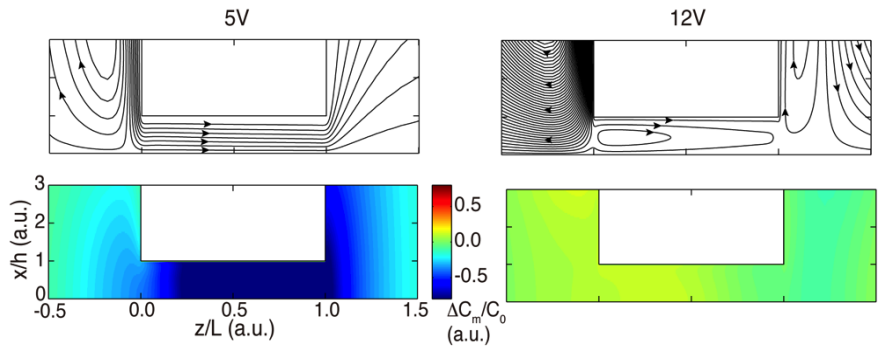

*Fig. 2. Simulated fluid flow lines (top) and the change of ion concentration induced by the fluid flow, $\Delta C_m$ (bottom) in the channel region. Two $V_d$ biases, 5V and 12V, are examined representing conductance suppression and restoration, respectively.*

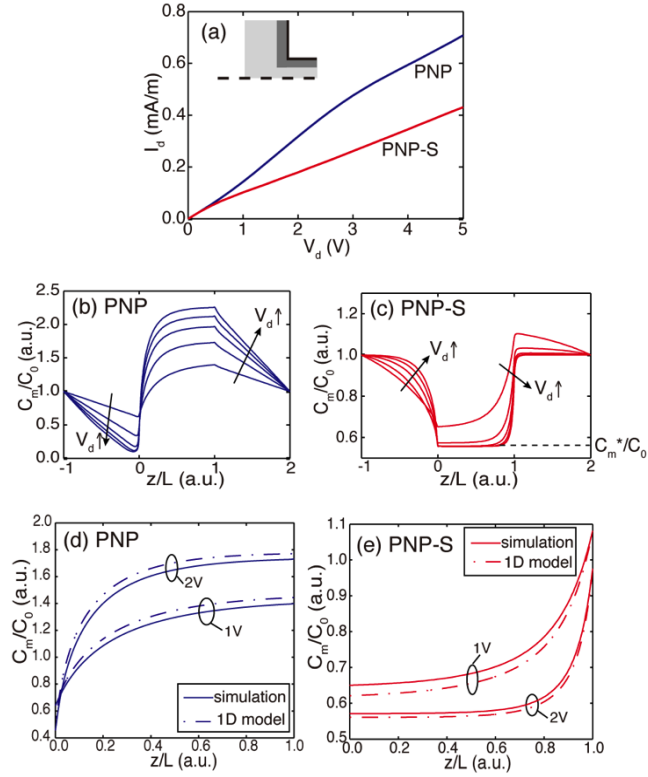

*Fig. 3.(a) Current-voltage curves simulated from PNP and PNP-S models for a long-channel structure. The inset schematically shows the unipolar (dark gray) and ambipolar (light gray) portions of the channel; (b)&(c) Ion concentration profiles along the central line ($x = 0$) from PNP (b) and PNP-S (c) simulations. The $V_d$ bias varies from 1V to 5V; (d)&(e) Ion concentration profiles of the channel portion ($0 \leq z \leq L$) from both numerical simulations and the 1-D model, without (d) and with (e) the account of EOF, respectively.*

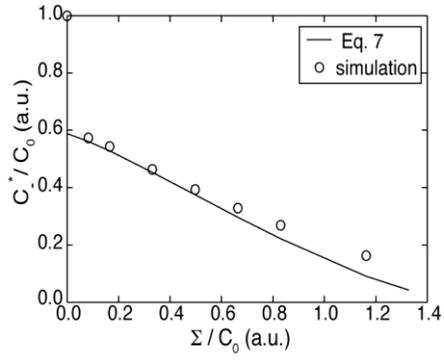

*Fig. 4. The limiting ion concentration, $C_-^*$, as a function of the surface charge, $\Sigma \equiv \sigma/h$, for the long-channel structure. Both results from PNP-S simulations and calculations based on Eq. 7 are shown.*

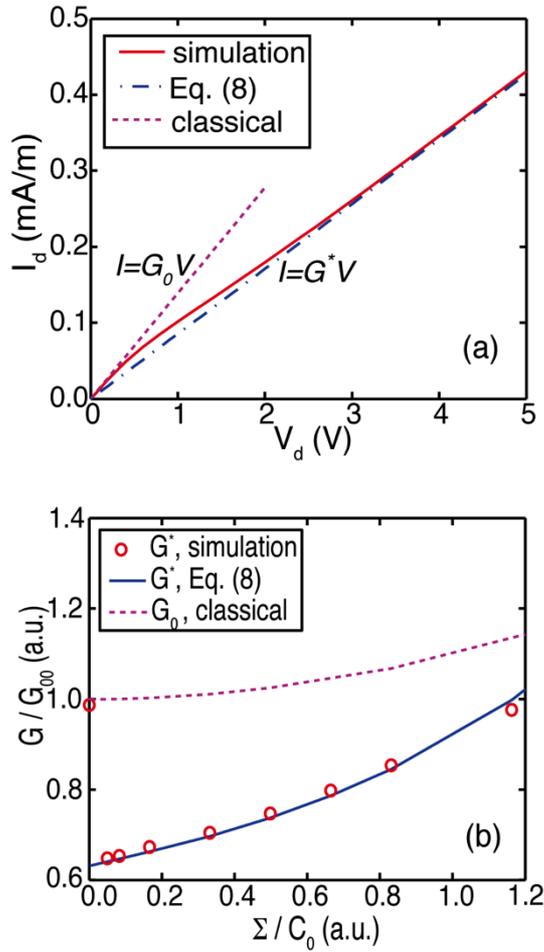

*Fig. 5.(a) Numerically simulated current-voltage curve using the PNP-S model, re-plotted from Fig. 3(a). The two asymptotes are based on the classical conductance [26], $G_0$, and the limiting conductance of Eq. (8), $G^*$, respectively; (b) The limiting conductance as functions of the surface charge $\Sigma$, from both PNP-S simulations and calculations based on Eq. (8). The classical conductance is also plotted as a reference. All values are normalized against the classical conductance at zero surface charge, $G_{00}$.*

*Corresponding author: yliu137@zju.edu.cn